\documentclass[11pt,twoside]{article}
\usepackage{asp2010}

\resetcounters

\begin{document}

\title{Investigations of a new eclipsing cataclysmic variable HBHA\,4705-03}
\author{D.G. Yakin,$^1$, V.F. Suleimanov,$^{2,1}$, V.V. Shimansky$^1$, V.V. Vlasyuk$^3$ 
and \\O.I. Spiridonova$^3$
\affil{$^1$Kazan (Volga region) Federal University, Kremlevskaya str. 18, 42008 Kazan, Russia}
\affil{$^2$Insitute for Astronomy and Astrophysics, Kepler Center for Astro and Particle Physics, Eberhard Karls University, Sand 1, 72076 T\"ubingen, Germany}
\affil{$^3$Special Astrophysical Observatory, 369167 N. Arkhyz, Russia}
}

\begin{abstract}
Results of photometric and spectroscopic investigations of the recently
discovered eclipsing cataclysmic variable star HBHA\,4705-03 are
presented.
The emission spectra of the system show broad  hydrogen and
helium emission lines. The bright spots with an approximately zero velocity components are found  
in the Doppler maps for the hydrogen and ionized helium lines. The disc structure is more prominent in the maps
for the neutral helium lines.
The masses of the components ($M_{\rm WD} = 0.54 \pm 0.10 M_{\odot}$ and
$M_{\rm RD} = 0.45 \pm 0.05~ M_{\odot}$), and the orbit inclination  
($i = 71.^{\circ}8 \pm 0.^{\circ}7$) were estimated  using the radial velocity light curve and the eclipse width.  
The modeling of the light curve allows us to evaluate the bright 
spot parameters and the mass accretion rate ($\dot  M \approx 2 \cdot 10^{17}$~g~s$^{-1}$).
\end{abstract}

\section{Introduction and Observations}

For the  first time HBHA 4705-03 ($\alpha_{2000}=22^{h}16^{m}50^{s}.8$, $\delta_{2000}=+46^{\circ}46.6$) 
was found as an object with $H_{\alpha}$ emission line \citep{catHA}, and
recently identified as an eclipsing cataclysmic variable (CV) \citep{KK:06}
with the orbital period 0.$^d$1718(1).

\begin{figure}
  \includegraphics[height=.25\textheight]{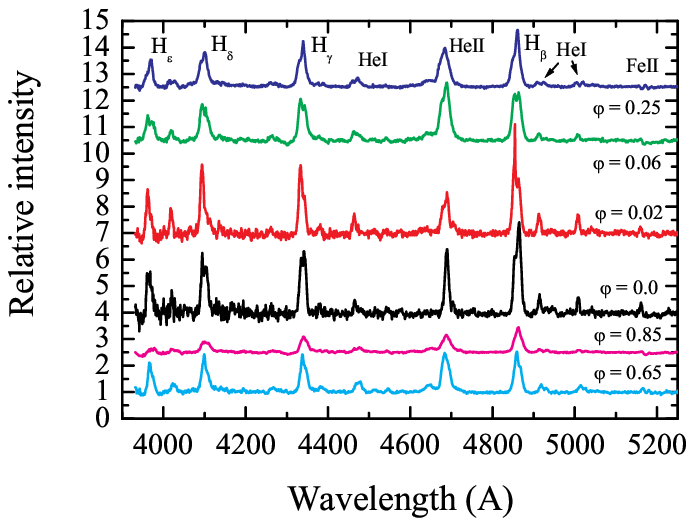}
  \includegraphics[height=.25\textheight]{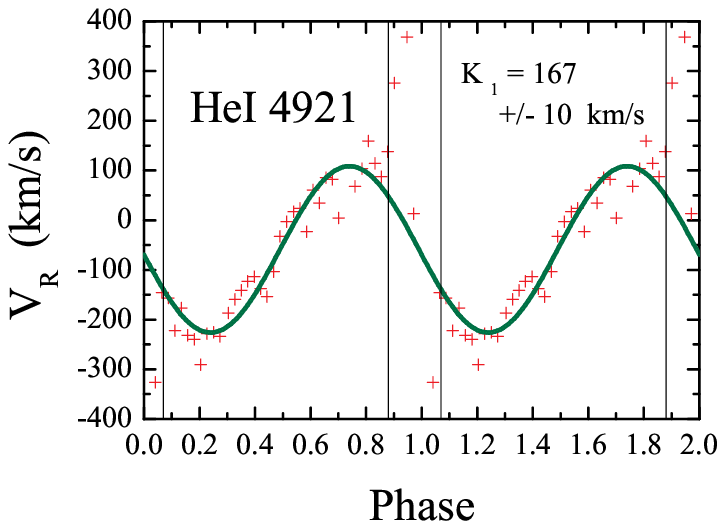}
  \caption{ \label{ykfig1}
{\it Left:} Normalized observed spectra in six orbital phases. {\it Right:} Dependence of the radial velocity on the orbital phase, obtained using HeI\,4921 emission line (21.07.07). The sine which fits the
radial velocity curve in 0.1 -- 0.9 phase range are shown by the solid curve.}
\end{figure}

The photometric observations of this close binary were performed on
August, 15-16 and 26-27, 2006 (V band) and on August, 16-17, 2006 (B band), with the  1--meter 
telescope  Zeiss1000 at the Special Astrophysical Observatory (Russia)
with an exposure time of 120 sec.
The total observation
time  was about 11 hours and the observations were calibrated against SDSS standard stars.
The example of the obtained light curves is shown in Fig.\ref{ykfig4}. These observations  give the ephemeris 
of the system eclipse ( HJD = 2453974.491(2) + 0.$^d$171814(27)$\times E$).
Spectroscopic observations of HBHA 4705-03 were carried out on August, 30-31, 2006 and July, 21-22, 2007,
by the 6--meter telescope BTA of the Special Astrophysical Observatory with
the SCORPIO focal reducer \citep{AM:05}, which gives a  $\Delta\lambda$ = 5.0
\AA~ resolution in the wavelength region  $\Delta \lambda$ 3900--5700
\AA. 16 subsequent spectra  with the same exposure time of 300 s
and the signal-to noise ratio  $S/N \approx$ 55 - 65 were obtained. 
Examples of the obtained spectra are shown in Fig.\ref{ykfig1}.
The orbital phases are counted from the photometric minimum.

\begin{figure}
\begin{center}
  \includegraphics[height=.38\textheight]{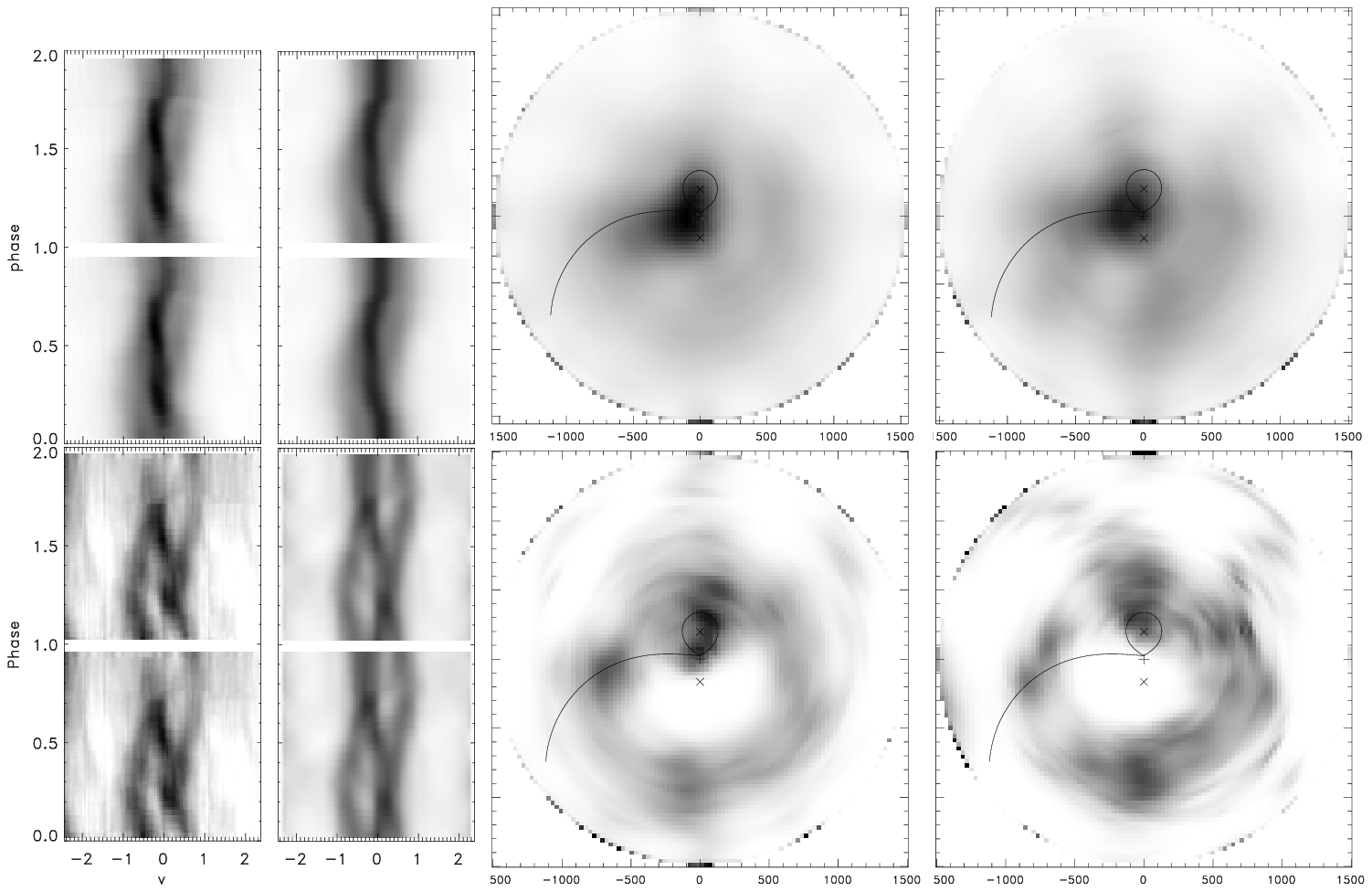}
\includegraphics[bb= 183 764 410 787,clip,angle=90,width=0.05\columnwidth]{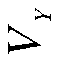}
\includegraphics[bb= 129 759 390 787,clip,angle=0,width=0.5\columnwidth]{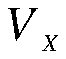}
  \caption{ \label{ykfig2}
Two left columns of the panels: trailed spectra of H$_{\beta}$ (top panels) and HeI 4921
(bottom panels), observed (left panels) and restored  from the Doppler maps (right panels). Velocities in the 
X-axes are in 1000 km s$^{-1}$ .
Two left columns of the panels: Doppler maps for the same lines for two nights (30.08.06 and 21.07.07).}
\end{center}
\end{figure}

\section {Data Analysis}

The most prominent details in the observed spectra of HBHA 4705-03 are the broad single
peaked emission lines of
hydrogen as well as  neutral and ionized helium (Fig.~\ref{ykfig1}, left panel). The line profiles are strongly
changing  with the orbital phase, especially at the moment of the eclipse, when they become double-peaked.

The radial velocity of the white dwarf was measured using the emission hydrogen lines by Shafter's method \citep{sh:83}. Dependence of the radial velocity on the orbital phase for  HeI\,4921 is
shown in Fig.~\ref{ykfig1} (right panel). The radial velocity curves were fitted by sine in the phase
range 0.1 -- 0-9.

Doppler maps of the system  for seven spectral emission lines 
were created using
Spruit's code \citep{sp:98}. The trailed spectrogram for H$_{\beta}$ and corresponding Doppler map ,
and the same for HeI\,4921are shown in Fig.~\ref{ykfig2}. It is clear from the Doppler maps that there 
is a bright spot at $L_1$ and an extended disc structure. 

\begin{figure}
  \includegraphics[height=.27\textheight]{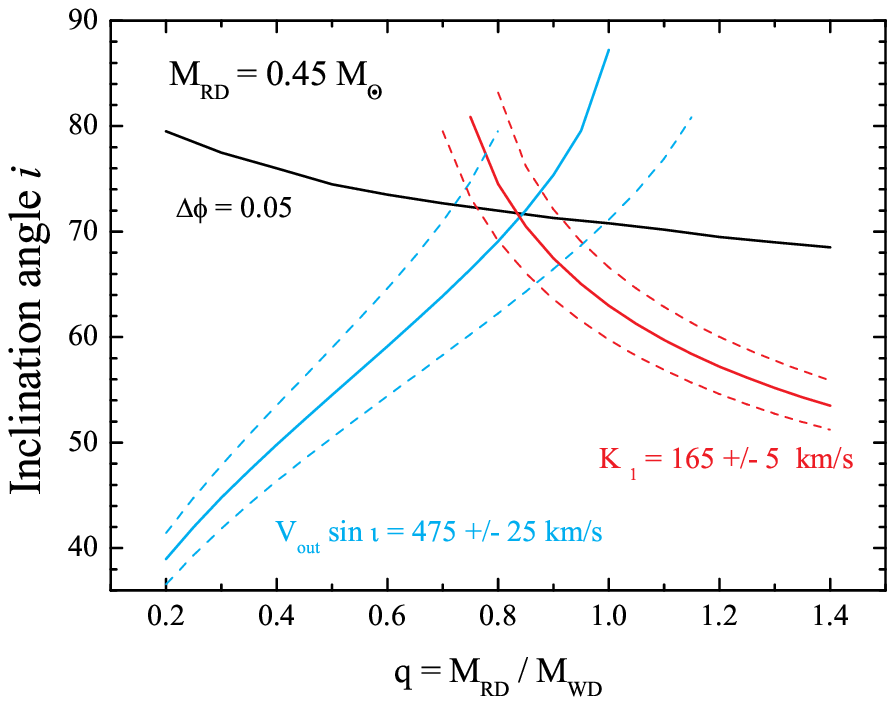}
  \includegraphics[height=.24\textheight]{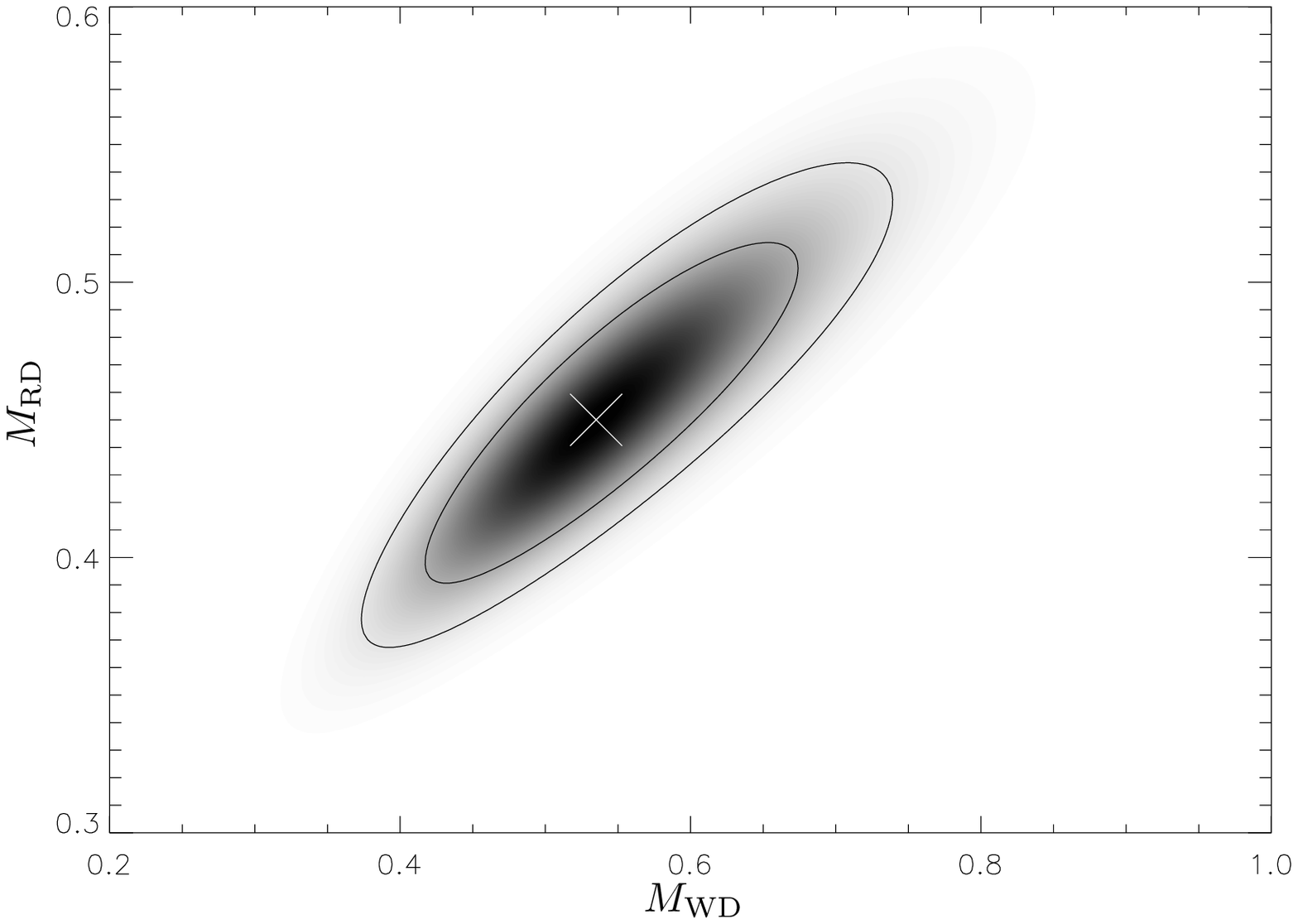}
  \caption{ \label{ykfig3}
{\it Left:}  The relationships between the mass ratio $q$ and the inclination angle $i$ 
 calculated for $M_{\rm RD} = 0.45 M_{\odot}$ and for the fixed values of the observed parameters.
 {\it Right:} Distribution of the total probability that the observed data correspond to given 
$M_{\rm WD} - M_{\rm RD}$ pair. 1- and 3- sigma contours are also shown.}
\end{figure}

\section {Estimation of system parameters}

 
The eclipse width ($\Delta \varphi \approx$   0.05) provides us the relation between the inclination angle
 $i$ and the mass ratio $q = M_{\rm RD}/ M_{\rm WD}$ \citep{horn:85}. 

The amplitude of the radial velocity $K_1$ = 165 $\pm$ 5 km s$^{-1}$ and the Kepler velocity at the outer 
disc radius $V_{\rm out} \sin i$ = 475 $\pm$ 25 km s$^{-1}$ determined  from the half of the distance 
between emission line peaks \citep[see details in][]{Yk:10,Yk:11}
 give two additional relations between $M_{\rm RD}$, $i$ and $q$.
We take  both values from the measurement of HeI lines because Balmer and HeII lines are distorted by the
strong component which arises at the inner Lagrange point. Using this information we
calculated the allowed regions in the $i$--$q$ and $M_{\rm WD}$ -- $M_{\rm RD}$ planes shown in
Fig.~\ref{ykfig3}. Finally, we have $M_{\rm WD} =
0.54 \pm 0.1 M_{\odot}$, $M_{\rm RD} = 0.45 \pm 0.05~ M_{\odot}$, and $i = 71.^{\circ}8 \pm 0.^{\circ}7$. 

\begin{figure}
  \includegraphics[height=.28\textheight]{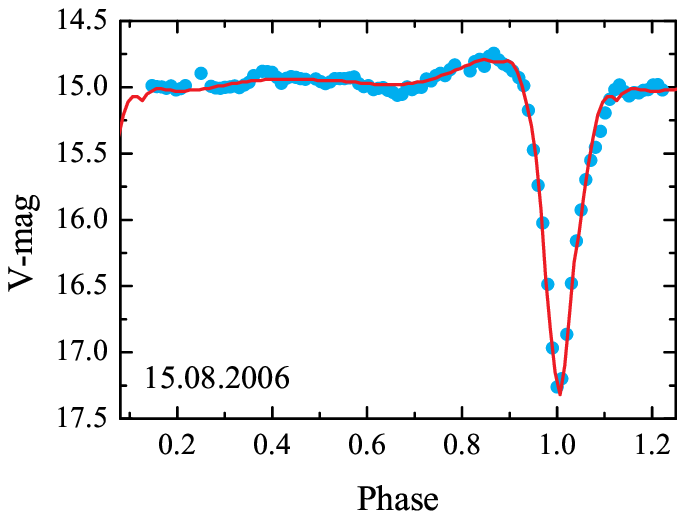}
  \includegraphics[height=.25\textheight]{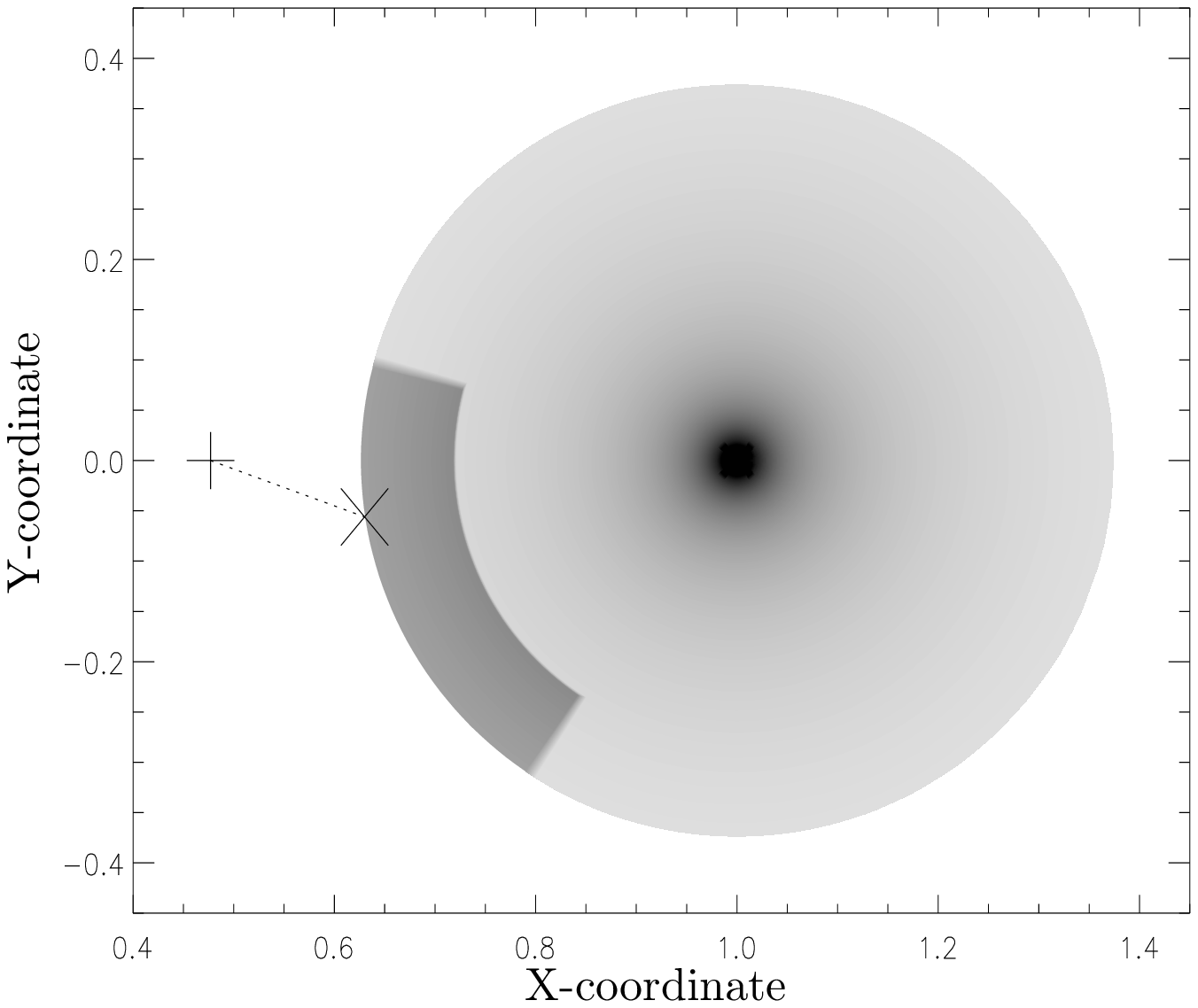}
  \caption{ \label{ykfig4}
{\it Left:}  The observed light curve together with the best model fit.
 {\it Right:} The model of an accretion disk with the bright spot. 
The red dwarf is on the left, at $X<0.4$. The brightness of the disk regions corresponds to the temperature.
The accretion stream trajectory is shown by the dashed line.}
\end{figure}

We modeled the observed light curve using Stupalov's code \citep[see details in][]{Yk:11} with the fixed system parameters mentioned 
above. As a result we estimated the bright spot  geometry and the temperature ($\approx$ 16000 K) together
 with the mass accretion rate $\dot M \approx 2 \cdot 10^{17}$ g s$^{-1}$. The results are shown in 
 Fig.\ref{ykfig4}.  The mass accretion rate and the mass of the secondary are in a good agreement 
 with the expected values estimated from the empirical relation $M_{\rm RD}$ -- $P_{\rm orb}$ \citep{knig:06}
  and the theoretical calculations \citep{rap:01}.
 
\acknowledgements This work is supported by the
Russian Foundation for Basic Research (grant  12-02-97006-r-povolzhe-a) and the DFG SFB\,/\,Transregio 7 ``Gravitational Wave Astronomy'' (V.S.).

\bibliography{suleimanov_poster}

\begin{thebibliography}{}
\expandafter\ifx\csname natexlab\endcsname\relax\def\natexlab#1{#1}\fi
\expandafter\ifx\csname url\endcsname\relax
  \def\url#1{\texttt{#1}}\fi
\expandafter\ifx\csname urlprefix\endcsname\relax\def\urlprefix{URL }\fi
\providecommand{\eprint}[2][]{\url{#2}}

\bibitem[{{Afanasiev} \& {Moiseev}(2005)}]{AM:05}
{Afanasiev}, V.~L., \& {Moiseev}, A.~V. 2005, Astronomy Letters, 31, 194

\bibitem[{{Horne}(1985)}]{horn:85}
{Horne}, K. 1985, \mnras, 213, 129

\bibitem[{{Howell} et~al.(2001){Howell}, {Nelson}, \& {Rappaport}}]{rap:01}
{Howell}, S.~B., {Nelson}, L.~A., \& {Rappaport}, S. 2001, \apj, 550, 897

\bibitem[{{Knigge}(2006)}]{knig:06}
{Knigge}, C. 2006, \mnras, 373, 484

\bibitem[{{Kohoutek} \& {Wehmeyer}(1997)}]{catHA}
{Kohoutek}, L., \& {Wehmeyer}, R. 1997, Abhandlungen aus der Hamburger
  Sternwarte, 11

\bibitem[{{Korotkiy} \& {Kryachko}(2006)}]{KK:06}
{Korotkiy}, S.~A., \& {Kryachko}, T.~V. 2006. \eprint{vsnet-allert/8893}

\bibitem[{{Shafter}(1983)}]{sh:83}
{Shafter}, A.~W. 1983, \apj, 267, 222

\bibitem[{{Spruit}(1998)}]{sp:98}
{Spruit}, H.~C. 1998, ArXiv Astrophysics e-prints.
  \eprint{arXiv:astro-ph/9806141}

\bibitem[{{Yakin} et~al.(2011){Yakin}, {Suleimanov}, {Borisov}, {Shimanskii},
  \& {Bikmaev}}]{Yk:11}
{Yakin}, D.~G., {Suleimanov}, V.~F., {Borisov}, N.~V., {Shimanskii}, V.~V., \&
  {Bikmaev}, I.~F. 2011, Astronomy Letters, 37, 845

\bibitem[{{Yakin} et~al.(2010){Yakin}, {Suleimanov}, {Shimansky}, {Borisov},
  {Bikmaev}, \& {Sakhibullin}}]{Yk:10}
{Yakin}, D.~G., {Suleimanov}, V.~F., {Shimansky}, V.~V., {Borisov}, N.~V.,
  {Bikmaev}, I.~F., \& {Sakhibullin}, N.~A. 2010, in American Institute of
  Physics Conference Series, edited by K.~{Werner}, \& T.~{Rauch}, vol. 1273 of
  American Institute of Physics Conference Series, 346

\end{thebibliography}
\end{document}